\documentstyle[preprint,aps]{revtex}

\begin{document}  
\draft 
  
\title{ Breached superfluidity
of fermionic atoms in magnetic field.}  
\author{G.M.Genkin$^*$.}
\address{ Physics Department and
Center for Polymer Studies, Boston University ,
              Boston, MA 02215.}

\maketitle 
                 
\begin{abstract}

  We 
derived the energy gap 
of a breached pairing superfluidity phase of 
 fermionic atoms in
an external magnetic field in
 Feshbach
resonance experiments which is determined by 
 the magnetic - field detuning from the Feshbach resonance.
We show that a
BCS
 superfluid state exists only for the magnetic - field
detuning  smaller than  
one critical, and this critical magnetic - field detuning is determined
by the equality of the Zeeman energy splitting for the magnetic - field
detuning to the energy gap $ \Delta_0 $.

\end{abstract}  

%\maketitle 
  
   The 
trapping and cooling of gases with Fermi statistics has become one of the
central areas of research within the field of ultracold atomic
gases. Much
progress has been made in the achievement
of degenerate regimes of trapped atomic Fermi gases [ 1 - 5]. The major
goals of studies of these systems is to observe a transition to a
paired - fermion superfluid state.
 There has been considerable interest
in achieving superfluidity in an ultracold trapped Fermi gas in which a
Feshbach resonance is used to tune the interatomic attraction by
variation of a magnetic field. The interactions which drive the pairing
in
these gases can be controlled using a Feshbach resonance, in which a
molecular level is Zeeman tuned through zero binding energy using an
external
magnetic field. Via a Feshbach resonance it is possible to tune the
strength
and the sign of the effective interaction between particles. In result,
magnetic - field Feshbach resonances provide the means for controlling
the
strength of cold atom interactions, characterized $ s $ - wave
scattering
length $ a $, as well as whether they are
effectively repulsive ( $ a > 0 $) or attractive ( $ a < 0 $ ).
Therefore, the tunability of interactions in fermionic atoms provides a
unique
possibility to explore the Bose - Einstein condensate to
Bardeen - Cooper -Schriever ( BEC - BCS ) crossover [ 6 - 8 ], an
intriguing interplay between the superfluidity of bosons and Cooper
pairing of fermions. A Feshbach resonance offers the unique possibility
to
study the crossover between situations governed by Bose - Einstein and
Fermi - Dirac statistics. When the scattering length $ a $ is positive
the
atoms pair in a bound molecular state and these bosonic dimers can form
a
Bose - Einstein condensate; when $ a $ is negative, one expects the
well - known BCS model for superconductivity to be valid.

On the other hand, recently has been considerable interest
in the study of asymmetric fermionic systems and the possibility of new
form of superfluidity in this systems.
An interesting new phase in Fermi matter, termed as breached pairing
state,
has been recently predicted by Liu and Wilczek [9]. This pairing
phenomenon
gives rise to a superfluidity having superfluid and normal Fermi
liquid components simultaneously. The pairing between fermions with
different
Fermi momenta in asymmetric fermion systems produces a state with
coexisting
superfluid and normal fluids. Possible realization of this phase was
considered for different Fermi systems (  an analogous
state for superconductivity in conventional superconductors in a strong
spin - exchange field was predicted many years ago by Sarma [10], in
ultracold fermionic atom systems composed of two particle species with
different densities and unequal masses [11], in quantum chromodynamics
in the context of color superconductivity [12] ).

  We will consider the breached superfluidity of cold fermionic atoms in
an external
 magnetic field under conditions of Feshbach resonance experiments.
Usually, these experiments are initiated by preparing atoms in a mixture
of states with different spin projections ( for example, spin - up and
spin - down ) with equal populations. We show that  the
energy gap parameter of a breached pairing superfluidity state
in Feshbach resonance experiments
 is
determined
by the magnetic - field detuning $ \Delta B $ from the Feshbach
resonance.
We show that a
BCS
 superfluid state (SF)
  exists only for the magnetic - field detuning smaller than one
  critical, and this critical magnetic - field detuning is determined
  by the equality of the Zeeman energy splitting for the magnetic -
  field
detuning to
the energy gap $ \Delta_0 $ of fermionic atoms without a magnetic field.
A SF state is lost for $ \Delta B $ larger than a critical.
For example, for $ ^{6} Li $ atoms this critical magnetic - field
detuning $ \Delta B_{cr} $ is order of $ 200 mG $, and for
$ \Delta B > 200 mG $ a SF state is lost. Note that the destroying a
superfluid state for the magnetic - field detuning $ \Delta B $
larger than $ \Delta B_{cr} $
 corresponds to a well - known dichotomy between superconductivity
and ferromagnetism. The strong field destroys the
superfluid state when the field is strong enough to break Cooper pairs, and
an external magnetic field provides the
pair - breaking mechanism.

  We consider an uniform gas of Fermi atoms with two hyperfine ( spin -
  up  $ \uparrow $
and spin - down  $ \downarrow $ ) states in an external
 magnetic field. Our
  starting model can be described by a Hamiltonian
$$
H - \mu_{\uparrow} N_{\uparrow} - \mu_{\downarrow} N_{\downarrow}
  = \sum_{\bf p}[( \varepsilon_p - \mu_{\uparrow} 
- \mu_{mag} B ) a^{+}_{{\bf p} \uparrow}
a_{{\bf p} \uparrow}  +
 $$
 $$ 
 ( \varepsilon_p - \mu_{\downarrow} + \mu_{mag} B) a^{+}_{{\bf p} \downarrow}
a_{{\bf p} \downarrow}] - \frac{g}{V} \sum_{{\bf p},{\bf q}} a^{+}_{{\bf
    p} \uparrow } a^{+}_{{ -\bf p} \downarrow} a_{{ -\bf q} \downarrow}
a_{{ \bf q} \uparrow },  \eqno (1)
   $$
with the coupling constant $ g = \frac{ 4 \pi h^2 |a|}{m} $, volume $ V $.
Here a coupling constant corresponds the attractive ( $ a < 0 $ )
pairing
interaction; $ a_{\bf p \sigma} ( a^{+}_{\bf p \sigma}) $ represent
the annihilation (creation) operators of a Fermi atom with the kinetic energy
$ \varepsilon_p = \frac{p^2}{2 m} $, and  $ \mu $ is the chemical
potential.
The Zeeman energy in an external magnetic field $ \bf B $ is
$ -\beta \bf{\sigma} \bf B $, and corresponding terms for spins $\uparrow $
and $ \downarrow $ are $ \pm \mu_{mag} B $ where $ \mu_{mag} $ is the
atomic magnetic moment.
In general, $ \mu_{\uparrow}$ and $ \mu_{\downarrow}$ are not equal.
Usually, in Feshbach resonance experiments
the chemical potentials of the states are determined by experimental
densities. In these experiments there is using
the radiofrequency
driving
the Zeeman transition $ \Delta_{Zee}^{(0)}$
 between spin states $\uparrow $ and $ \downarrow $, and 
these states are preparing with equal populations. Here
$ \Delta_{Zee}^{(0)} $ is the Zeeman energy splitting for the magnetic
field $ B_0 $, i. e. $ \Delta_{Zee}^{(0)} = 2 \mu_{mag} B_0 $, and $ B_0
$ is the magnetic field in the vicinity of the Feshbach resonance.
Due to equal populations of spin states $ \uparrow $ and $ \downarrow $
 we will assume
  $$
\mu_{\downarrow} = \mu_{\uparrow} + \Delta_{Zee}^{(0)}.       \eqno (2)
  $$
It is correspond to well - known statement [13] that for a spin in a
magnetic
field we may assume that the chemical potential is equal to
$ \mu_{\uparrow} + \mu_{mag} B $ for atoms with the spin projection
$ \uparrow $ along the field and to $ \mu_{\downarrow} - \mu_{mag} B $
for atoms with the spin projection $ \downarrow $ opposite to the field.
 In the Hamiltonian (Eq.(1))
we introduce the standard canonical transformation to the Bogolyubov
 quasiparticles
  $$
a_{{\bf p} \uparrow} = u_p b_{{\bf p} \uparrow} +
 v_p b^{+}_{-{\bf p} \downarrow},
  $$
  $$
a_{{\bf p} \downarrow} = u_p b_{{\bf p} \downarrow} -
 v_p b^{+}_{-{\bf p} \uparrow},        \eqno (3)
  $$
where the
coefficients $ u_p $ and $ v_p $ are real, depend only on $ |p| $. They
are
chosen from the condition [14] that the energy $ E $ of the system has a
minimum for a given entropy. We have
  $$
E  = \sum_{\bf p} ( \varepsilon_p - \mu_{\uparrow}
 - \mu_{mag} B) [u_p^2 n_{{\bf p} \uparrow}
  + v_p^2 ( 1 - n_{{\bf p} \downarrow})] +
  $$
  $$
  ( \varepsilon_p - \mu_{\downarrow} +
  \mu_{mag} B) [u_p^2 n_{{\bf p} \downarrow}
  + v_p^2 ( 1 - n_{{\bf p} \uparrow})] -
 \frac{g}{V}[\sum_{\bf p} u_p v_p ( 1 - n_{{\bf p} \uparrow} -
 n_{{\bf p} \downarrow})]^2.           \eqno (4)
 $$
Varying this expression with respect to the parameter $ u_p $ and using
the
relation $ u_p^2 + v_p^2 = 1 $, which the transformation coefficients
must be satisfy, the condition for a minimum is
$\frac{\delta E}{\delta u_p} = 0 $. Using this condition for Eq.(4)
we have the standard [14]
equation for the energy gap
$$
\frac{g}{2 V} \sum_{\bf p} \frac{( 1 - n_{{\bf p} \uparrow} -
 n_{{\bf p} \downarrow})}{ \sqrt{ \Delta^2 + \eta_p^2}} = 1,    \eqno (5)
$$
where the energy gap
  $$
\Delta =\frac{g}{V} \sum_{\bf p} u_p v_p ( 1 - n_{{\bf p} \uparrow} -
 n_{{\bf p} \downarrow})        \eqno (6a)
$$
and
$$
u_p^2 = \frac{1}{2} ( 1 + \frac{\varepsilon_p^{+}}{\sqrt { \Delta^2 +
 \varepsilon_p^{+2}}}),
v_p^2 = \frac{1}{2} ( 1 - \frac{ \varepsilon_p^{+}}{ \sqrt{ \Delta^2 +
 \varepsilon_p^{+2}}}),    \eqno(6b)
$$
where
$$
2 \varepsilon_p^{+} = ( \varepsilon_p - \mu_{\uparrow} - \mu_{mag} B ) +
(\varepsilon_p - \mu_{\downarrow} + \mu_{mag} B ) = 2 \varepsilon_p -
( \mu_{\uparrow} + \mu_{\downarrow} ),     \eqno (6c)
$$ 
and the quasiparticle occupation numbers $ n_{{\bf p} \alpha} =
 b_{{\bf p}\alpha}^{+} b_{{\bf p}\alpha} $.
Note that due to the standard canonical transformation ( Eq.(3))
the Cooper pair has the zero summary momentum, therefore, we have
an uniform energy gap $ \Delta $ ( Eq.(5)).
The energy of the elementary excitations $ E_{\uparrow}(p),
E_{\downarrow}(p) $
 can be find [14] from the
change of the energy $ E $ of the system when the quasiparticle
occupation
numbers are changing, i. e. by varying $ E $ with respect to
$  n_{{\bf p} \uparrow} $
and $ n_{{\bf p} \downarrow} $. Therefore, start from the equation
$$
\delta E = \sum_{\bf p} [ E_{\uparrow}(p) \delta n_{{\bf p} \uparrow}
+ E_{\downarrow}(p) \delta n_{{\bf p} \downarrow}],    
$$
we have
two branches of quasiparticle excitations with
the spectra
$$
E_{\uparrow}(p) = \frac{ \delta E}{\delta n_{{\bf p}
    \uparrow}} = \sqrt{\varepsilon_p^{+2} + \Delta^2} -
\frac{1}{2} \delta \Delta_{Zee} ,     
$$
$$
E_{\downarrow}(p) = \frac{ \delta E}{\delta n_{{\bf p}
    \downarrow}} = \sqrt{\varepsilon_p^{+2} + \Delta^2} + 
\frac{1}{2}\delta \Delta_{Zee} ,     
$$
$$
\delta \Delta_{Zee} = \Delta_{Zee} - \Delta_{Zee}^{(0)} =
2 \mu_{mag} ( B - B_0).      \eqno (7)
$$
 The quasiparticle occupation numbers satisfy Fermi - Dirac statistics.

For $ | \delta \Delta_{Zee} | < 2 \Delta $ we have a BCS superfluidity
with both gapped excitations ( Eqs.(7)). Meanwhile, for 
$ | \delta \Delta_{Zee} | > 2 \Delta $ one branch of quasiparticle
excitations ( for $ \Delta B = B - B_0 > 0 $ the branch $ E_{\uparrow}(p) $,
for $ \Delta B < 0 $ the branch $ E_{\downarrow}(p) $ )
 may be negative for momenta
$ p_1 \leq p \leq p_2 $ where
  $$
\frac{ p_{1,2}^2}{ 2 m} =\frac{ \mu_{\uparrow} + \mu_{\downarrow} }{2}
\pm \sqrt{ \frac{1}{4} \delta \Delta_{Zee}^2 - \Delta^2 },     \eqno (8)
  $$
and we have a breached pairing superfluidity with one gapped excitation 
branch and one gapless.In this case for finding the energy gap
$ \Delta $ we will follow the method of Ref.[13], Chapter 21.2.
  In the BCS gap equation ( Eq.(5))
using cutting off [14] the logarithmic integral
at same $ \eta =\bar{\epsilon} ( \bar{\epsilon} $
is the ultraviolet cutoff )
 we
have the well - known result for the energy gap $ \Delta_0(g) $ for
$ T = 0 $ and
 $ B = 0 $
$$
ln \frac{\bar{\epsilon}}{\Delta_0(g)} = \frac{  2 \pi^2 h^3}
{ g m p_F}.     \eqno (9a)
$$
 For $ B \neq 0 $, in general, the energy
gap
depends on the magnetic field $ \Delta(g,B)$.
If
$ \Delta(g,B) < \frac{1}{2} | \delta \Delta_{Zee}B | $ there is a nonzero 
minimal value in the integral ( Eq.(5))
$ \varepsilon_{p (min)}^{+} = \sqrt{\frac{1}{4} \delta \Delta_{Zee}^2 - \Delta^2} > 0 $, and
using the table integral $ \int\frac{dx}{\sqrt{x^2 + a^2}} = ln | x +
 \sqrt{ x^2 + a^2} | $
for $ \Delta < \mu_{mag}| \Delta B| $
we have in this case
$$
ln \frac{\bar{\epsilon}}{ \frac{1}{2} |\delta \Delta_{Zee}|  +
 \sqrt{ \frac{1}{4} \delta \Delta_{Zee}^2 - 
\Delta^2(g,B)}} = \frac { 2 \pi^2 h^3}{g m p_F}.     \eqno (9b)
$$
By comparing Eqs.(9) the energy gap $ \Delta(g,B) $ is determined by the
equation
$$
\frac{ |\delta \Delta_{Zee}|}{2} + \sqrt{ \frac{1}{4} \delta \Delta_{Zee}^2 - 
\Delta^2(g,B)} = \Delta_0(g),      \eqno (10a)
$$
or
$$
\Delta(g,B) = \sqrt{\Delta_0(g) (|\delta \Delta_{Zee}| - \Delta_0(g) )},
$$  
$$
 \delta \Delta_{Zee} = 2 \mu_{mag} \Delta B, \Delta B = B - B_0.      \eqno (10b)
$$
We derived the energy gap $ \Delta(g,B)$ of the
breached pairing superfluidity phase
  for
fermions in an external magnetic field under actual conditions of Feshbach
resonance
experiments. This  breached pairing superfluidity phase
  exists in the range $ 0 \leq \Delta (g,B) \leq \Delta_0(g) $ for
$ \Delta_0(g) \leq \delta\Delta_{Zee} \leq 2 \Delta_0(g) $.
 The equation (10a) has a threshold at the
value
$ | \delta \Delta_{Zee}^{cr}| = \Delta_0(g) $ for which this equation first has
solutions  for $|\Delta B| \succeq \Delta B_{cr} $, and, accordingly, the energy gap
$ \Delta(g,B) $ has a solution  for   $ |\Delta B| \succeq \Delta B_{cr}$.
In result, we have a breached pairing superfluidity phase with the
critical
magnetic - field detuning from the Feshbach resonance $ \Delta B_{cr} $.

  Note that for fermions in an external  magnetic field for
 $ \mu_{\uparrow} = \mu_{\downarrow} $ ( without a radiofrequency
driving ) the energy gap parameter of the
breached pairing superfluidity phase
 is determined by $ \Delta _{Zee} = 2 \mu_{mag} B $
 instead $ \delta \Delta_{Zee} $
( Eq.(10b)), and in this case for the conventional superconductors in a
 strong
exchange field $ B_{ex} $ there is a Sarma unstable  phase [10] with
 $ \Delta _{Zee}^{ex} = 2 \mu_{mag} B_{ex} $ in Eq.(10b).

 We consider the thermodynamic properties of Fermi atoms with two
 hyperfine
states 
in the breached pairing superfluidity phase
(for the magnetic - field detuning
  $  |\Delta B| \succeq \Delta B_{cr} $).
 In calculations it is convenient
to start from the thermodynamic potential $ \Omega $, and
the condensation energy
( the difference
between the thermodynamic potential $ \Omega_s $ in the superfluid state
 and
the value in the normal state $ \Omega_n $ at the same temperature [14])
 is given by ( at $ T = 0 $)
$$
 \Omega_s - \Omega_n = -\int_{0}^{g} \frac{\Delta^2(g_1)}{g_1} dg_1.
 \eqno (11)
$$
Changing in Eq.(11)
from integration over $ dg_1 $ to that over
$ d\Delta $, and using Eq.(10b) we obtain the difference between the
ground - state energies of the superfluid $ E_s $ and normal $ E_n $
states for magnetic - field detuning 
 $ 2 \mu_{mag}|\Delta B| \succeq \Delta_0 $, or
$|\delta \Delta_{Zee}| \succeq \Delta_0 $
 $$
E_s - E_n = \frac{ m p_F}{ 4 \pi^2 h^3} (\delta \Delta_{Zee} - \Delta_0 )^2.
\eqno (12a)
$$
The positive sign of this difference $ E_s - E_n > 0 $
indicates that for
magnetic -  field detuning
 $|\delta \Delta_{Zee}| > \Delta_0 $ the normal state has a smaller
energy.
The condensation energy is positive, indicating that this
breached superfluidity phase
 is
unstable, and the normal state is energetically favored.
 For magnetic - field detuning 
 $ |\delta \Delta_{Zee}| < \Delta_0 $ 
 we have a BCS state. In result, our system passes from
a BCS state to an
unstable breached superfluidity
 state as the magnetic -  field detuning from the
Feshbach resonance
 $ |\Delta B | = | B - B_0 | $
 increases to a critical
value
$$
\Delta B_{cr} =  \frac{ \Delta_0}{ 2 \mu_{mag}},   \eqno (13a)
$$
i. e. a critical value of the 
Zeeman splitting
for the magnetic - field detuning
 $\delta \Delta_{Zee}^{cr} = 2 \mu_{mag} \Delta B_{cr}  $ is
$$
 \delta \Delta_{Zee}^{cr}   =  \Delta_0.    \eqno (13b)
$$
By means of Eq.(13) we have
   $$
\Delta B_{cr} = 2 \frac{\Delta_0}{\Delta_{Zee}^{(0)}} B_0.   \eqno (14)
    $$

  For example, for $ ^{6} Li $ atoms a Feshbach resonance is located
  [15]
at $ B_0 \simeq 850 G $, the Fermi energy [15]
 $ \varepsilon_F \simeq 1.1\mu K $, and in experiments  there is
driving the Zeeman transition $ \Delta_{Zee}^{(0)} $ between the $ |1/2,
  1/2> $ and $ |1/2,-1/2> $
states with [16] $ \Delta_{Zee}^{(0)}  = 76 MHz $ rf field.
In the present experiments $ \Delta_0/\varepsilon_F = 0.2 - 0.4 $.
For  $ \Delta_0/\varepsilon_F \simeq  0.4 $ we have
$ \Delta B_{cr} \simeq 200 mG $ for  $ ^{6} Li $ atoms, and for
$ \Delta B > 200 mG $ a SF state is lost.

 For fermions in an external magnetic field with $ \mu_{\uparrow} =
\mu_{\downarrow} $ a BCS superfluid state exists only for the magnetic field
$ B < B_{cr} $, and this
 critical magnetic field  $  B_{cr} $
 which destroys the SF state
  is
determined from the condition that the Zeeman energy splitting
 $ \Delta_{Zee}^{(0)} $ 
equal to
 the energy gap
$ \Delta_0 $
 without a magnetic field.
Note that, although, an external
  magnetic
field does not penetrate a bulk conventional superconductor,
the strength of the field required to overcome the energy gap, i.e. to
destroy superconductivity in conventional superconductors, is much more
  than
for atomic Fermi gases because the energy gap  $ \Delta_0^{sup} $  of
the conventional superconductors is much more than $ \Delta_0 $ for
atomic
Fermi gases ,
usually, $ \Delta_0^{sup}/ \Delta_0 > 10^6 $.

  In summary,
we show 
that for fermionic atoms 
in an external magnetic field
in Feshbach
resonance experiments there is a breached pairing superfluidity 
for the magnetic - field detuning from the Feshbach resonance larger
than one critical. We derived the energy gap parameter of this phase
which is determined by the magnetic - field detuning;
 this unstable phase exists in the certain range of
the
Zeeman energy splitting determinable by the energy gap $ \Delta_0 $.
We show that a BCS superfluid state exists only for the magnetic - field
detuning from the Feshbach resonance smaller than  
the critical value, and this critical magnetic - field detuning is determined
by the equality of the Zeeman energy splitting for the magnetic - field
detuning to the energy gap $ \Delta_0 $.

   I thank H. E. Stanley for encouragement, J. Borreguero for assistance.

\end{document}